\documentclass[draft]{agujournal2019}
\usepackage{url} 
\usepackage{lineno}
\usepackage[inline]{trackchanges} 
\usepackage{soul}
\usepackage{amsmath}
\usepackage{tabularray}
\draftfalse
\journalname{Space Physics}
\begin{document}

\newcommand{\bzimf}[0]{B_{z}^{\tiny{\text{IMF}}}}
\newcommand{\bzimfmin}[0]{(B_{z}^{\tiny{\text{IMF}}})_{\text{min}}}
\newcommand{\vx}[0]{V_{x}}
\newcommand{\vxmax}[0]{(V_{x})_{\text{max}}}
\newcommand{\gicmax}[0]{|\text{GIC}|_{\text{max}}}
\newcommand{\gicmaxub}[0]{|\text{GIC}|_{\text{max[0.80]}}}
\newcommand{\aicc}[0]{\text{AICc}}
\newcommand{\pseudoaicc}[0]{\text{pseudo-\aicc}}

%
%


\title{Empirical Relationship for Geomagnetically Induced Currents (GIC) and Solar Wind Conditions}

%
%




\authors{Dean Thomas\affil{1}, Lucy A. Wilkerson\affil{1}, Robert S. Weigel\affil{1}, John Haman\affil{2}, Edward J. Oughton\affil{1}}

\affiliation{1}{Space Weather Lab, Department of Physics and Astronomy, George Mason University, Fairfax, VA, USA}
\affiliation{2}{Institute for Defense Analyses, Alexandria, Virginia, USA}





\correspondingauthor{Dean Thomas}{dthoma6@gmu.edu}



\begin{keypoints}
    \item An empirical model is developed to estimate $\gicmax$ using solar wind drivers and local $\alpha$ and $\beta$ scaling factors.
    \item Model stability and robustness are validated via cross-validation and Cook's distance across nine major storms ($K_p \ge 8$).
    \item Statistical analysis of 39 historical storms shows that the upper 80\% bounds of $\gicmax$ are consistent with a lognormal distribution.
\end{keypoints}

%
%

%
%


\begin{abstract}
Geomagnetically induced currents (GICs) during extreme space weather events represent a critical hazard to modern infrastructure, including electrical power grids, telecommunication systems, pipelines, and railways. This study presents empirical models designed to provide rapid, site-specific estimates of the maximum GIC, $\gicmax$, expected during a storm. Utilizing data from 67 sites across nine major geomagnetic storms ($K_p \ge 8$) occurring between 2020 and 2025, we performed multivariate linear and quantile regressions incorporating solar wind parameters and local scaling factors. The solar wind parameters address storm intensity and are the minimum interplanetary magnetic field $z$-component, $\bzimfmin$, and the maximum magnitude of the solar wind velocity $x$-component, $\vxmax$. The scaling factors account for local differences due to geomagnetic latitude ($\alpha$) and ground conductivity ($\beta$).  A 6-parameter model was identified via best subset selection and the corrected Akaike Information Criterion (AICc) as the optimal predictor for mean $\gicmax$, yielding a Pearson correlation of $r = 0.73$ and a coefficient of determination of $R^2 = 0.54$. To examine peak GICs during storms, a 7-parameter quantile regression model was developed to estimate the upper 80\% bound of $\gicmax$.  Statistical analysis of 39 severe storms since 1995 shows that the $\gicmax$ upper bounds are consistent with a lognormal distribution. These relationships offer a computationally efficient alternative to complex physics-based simulations, enabling space weather practitioners to estimate site-specific GICs in high-level analyses.
\end{abstract}

\section*{Plain Language Summary}
Extreme solar storms can trigger ``geomagnetically induced currents" (GICs) that flow through the Earth's surface and into critical infrastructure, like power grids and telecommunication systems. These currents can cause equipment failure and widespread power outages. In this study, we develop a simplified mathematical tool to predict the maximum size of these currents. This tool uses the speed and magnetic strength of the solar wind, alongside local factors like latitude and ground conductivity. By analyzing nine major solar storms between 2020 and 2025, we created a model to estimate ``worst-case'' current levels at specific locations. Further analysis of 39 historical storms since 1995 confirmed that these extreme current levels follow a well-known statistical pattern. Our formula provides a faster, simpler way for scientists and engineers to assess risks during solar storms without needing to run complex and time-consuming computer simulations.


%
%

%


%
%
%
%

\section{Introduction}

Geomagnetically induced currents (GICs) represent a significant threat to modern infrastructure, affecting systems such as electrical power grids, telecommunication systems, pipelines, and railway signals \cite{ngwira2019introduction, boteler2024examination, love2025impact, pirjola2012geomagnetically}. Studies have shown that there are significant economic consequences from extreme space weather events.  \citeA{oughton2024major} examined potential impacts on the United States from a 250-year geomagnetic storm.  They estimated daily U.S. economic losses from transformer thermal heating of $\approx 2$ billion U.S. dollars per day, disrupting power for more than 6 million people and 155,000 businesses.

In power grids, GICs introduce a quasi-DC current in transformer windings, which can lead to half-cycle core saturation. This saturation leads to increased reactive power consumption, the generation of harmful harmonics, and localized heating that can damage insulation.  It can also lead to catastrophic equipment failure \cite{figueroa2025geomagnetic, kirkham2011geomagnetic, patel2016analysis}. While GICs are typically more intense at high geomagnetic latitudes, these risks are not confined to high-latitude regions.  For example, intense magnetic field variations and GIC peaks were observed in Latin America and New Zealand during the May 2024 ``Gannon" storm~\cite{caraballo2025impact,malone2026mana}.

Long-line telecommunications systems, including terrestrial systems and submarine cables, are also affected by GICs. For terrestrial systems, geomagnetically induced voltages between the grounding points of communication networks can cause system interference and damage \cite{boteler2024examination}. Similarly, for submarine cables, GICs affect the power feed to the repeaters (amplifiers placed at intervals along the cable to improve signal quality).  This is a concern for legacy cables and modern fiber-optic cables, which contain a copper conductor to carry power to the repeaters \cite{love2025impact}. 

In pipeline infrastructure, GICs can interfere with cathodic protection systems, potentially accelerating electrochemical corrosion of buried oil and gas lines \cite{pulkkinen2001recordings, boteler2013new}. GICs can also induce errors within railway warning lights, causing ``right side" and ``wrong side" failures. Right-side failures occur when unoccupied sections appear occupied, and wrong-side failures occur when occupied sections appear clear \cite{patterson2024modelling}. 

Since GICs can affect our technological infrastructure, tools to estimate the expected magnitude of GICs during severe solar storms are useful.  \citeA{wilkerson} recently published an article examining the Gannon storm, in which they assembled a large set of geomagnetic disturbance (GMD) data and identified empirical relationships for GICs in the contiguous United States.  They considered whether the magnitude of the maximum GIC, $\gicmax$, during the storm could be estimated using $\alpha$ and $\beta$, which are scaling factors that account for local differences due to geomagnetic latitude ($\alpha$) and ground conductivity ($\beta$) \cite{nerc2025reliability}. \citeA{wilkerson} constructed a regression model using:
\begin{equation}
    \gicmax = A_{\alpha\beta}\alpha\beta + B_{\alpha\beta} + \epsilon
\end{equation} 
where the $A_{\alpha\beta}$ and $B_{\alpha\beta}$ are the regression coefficients and $\epsilon$ is the regression noise term. A similar equation for electric fields is used in the \citeA{nerc2025reliability} geomagnetic disturbance standard.  For $\gicmax$, they found a Pearson correlation coefficient of 0.76 when using the Gannon storm data.

The coefficients of the regression models are expected to depend on storm intensity, so regression models based solely on the Gannon storm are not expected to apply to other storms.  In \citeA{wilkerson}, they performed the same analysis as above for the 10 October 2024 storm and also found a linear relationship, but with a regression coefficient for the $\alpha\beta$ term about one-half of that for the Gannon storm.

The analysis in this paper extends the \citeA{wilkerson} analysis to encompass nine storms across 67 GIC monitoring sites to determine whether an empirical relationship can be found that applies across a range of storms.  Our statistical analysis of the nine solar storms and the associated solar wind conditions identified a multivariate relationship between the observed $\gicmax$, the $z$-component of the interplanetary magnetic field, $\bzimf$, the $x$-component of the solar wind velocity, $\vx$, and the $\alpha$ and $\beta$ scaling factors for the GIC sites. This relationship addresses the intensity of the storms through $\bzimf$ and $\vx$ and the local conditions through $\alpha$ and $\beta$.  It provides a simple, computationally efficient method of predicting the maximum $|\text{GIC}|$ expected during a geomagnetic storm for a specified site.

\section{GIC, Solar Wind, and Scaling Factors}

The data analyzed in this paper focus on nine storms. The North American Electric Reliability Corporation (NERC) provided measured GIC from monitoring sites throughout the continental United States for all nine storms.  The Tennessee Valley Authority (TVA) provided additional data for the 2024 May storm (Gannon storm). The selected storms include all geomagnetic storms from 2020 through June 2025 listed on the NERC Electric Reliability Organization (ERO) data portal  \cite{NERC_ERO} with reported $K_p\geq8$. These storms were selected both for their intensity and data availability. Each measured GIC time series was inspected for significant errors, including a non-zero baseline; large, unphysical spikes; and periods of constant GIC values lasting longer than $2$~minutes. Additionally, any GIC time series with a cadence $\ge1$~minute or a gap in the recorded data longer than $5$~minutes was omitted.

Solar wind conditions and select geomagnetic indices during each event were retrieved from the OMNI dataset (\protect\citeA{OMNI2020_1hr}, \protect\citeA{OMNI2020_1min}) using their HAPI \protect\cite{Weigel2021} server. These parameters include $\bzimf$ and $\vx$ in GSE coordinates. 

Finally, the NERC $\alpha$ and $\beta$ scaling factors, originally presented in the NERC TPL--007 geomagnetic disturbance standard \cite{nerc2025reliability}, were calculated at each GIC monitor site. These scaling factors were originally developed to calculate a benchmark simulated electric field with components $x,y$ having amplitudes of
\begin{equation}
\begin{split}
    E_x &= E_{ox}\alpha(\lambda)\beta_x (\phi,\lambda) \\
    E_y &= E_{oy}\alpha(\lambda)\beta_y (\phi,\lambda)\thinspace,    
\end{split}
\label{eqn:nerc}
\end{equation}

\noindent
where $\lambda$ and $\phi$ are geomagnetic latitude and longitude respectively, $E_{0x}$ and $E_{0y}$ are 1-in-100-year peak geoelectric field amplitudes at a reference location, $\alpha=0.001e^{0.115\lambda}$ is a geomagnetic latitude scaling factor \protect\cite{nerc2025reliability}, and $\beta_x$ and $\beta_y$ are scaling factors that depend on the local ground conductivity as described in \citeA{Pulkkinen25, wilkerson}. The $\beta$ scaling factor at each site was calculated by linearly interpolating the $\beta$ values calculated at magnetotelluric sites onto a $100\times100$ grid spanning all geographic longitudes and latitudes of the GIC monitor sites.

The result is a database for 67 sites across nine storms.  The database contains the maximum GIC magnitude observed at each site ($\gicmax$), the minimum $z$-component of the interplanetary magnetic field ($\bzimfmin$), the maximum magnitude of the $x$-component of the solar wind velocity ($\vxmax$), and the site-specific $\alpha$ and $\beta$ scaling factors for each storm period. The data are provided in \ref{app:data}.  

\section{Statistical Analysis}
\label{section:statanalysis}

\subsection{Linear Fit}
\label{section:initial}

We analyze the GIC and solar wind data using the Python statsmodels package \cite{seabold2010statsmodels}. The analysis uses linear regression to model $\gicmax$ based on solar wind conditions and the $\alpha$ and $\beta$ scaling factors. We initially fit:
\begin{equation}
\gicmax = A_0 + A_1 \alpha + A_2 \beta + A_3 \bzimfmin + A_4 \vxmax + \epsilon
\label{eqn:initialfit}
\end{equation}
where $A_0$, $A_1$, $A_2$, $A_3$, and $A_4$ are the regression coefficients.  $\bzimf$ and $\vx$ are intended to capture the intensity of the storm through their impact on magnetic reconnection and compression of Earth's magnetosphere \cite{Baumjohann_2012}. And $\epsilon$ is the regression noise term.

We use $\alpha$ and $\beta$ as scaling factors in this analysis, but we acknowledge that using other factors, such as magnetotelluric transfer functions and network configuration, can provide more information on local conditions.  Nonetheless, $\alpha$ and $\beta$ are used in the NERC standards, and as we will see, they provide a reasonable fit to the measured data.

In the statistical analysis that follows, the parameters are standardized, in this case centered and scaled. Standardization is a common statistical practice when the parameters in linear regressions have different units, means ($\mu$), and standard deviations ($\sigma$) \cite{james2023statistical}.  Standardizing parameters reduces multicollinearity that can obscure statistically significant relationships between parameters. As we develop the regression analysis, we add higher-order terms, and multicollinearity concerns will become more important. The transformation to standardize a generic parameter, $X$, is $X_{std} = (X-\mu)/\sigma$.

\begin{figure}
    \centering
    \includegraphics[width=1.0\linewidth]{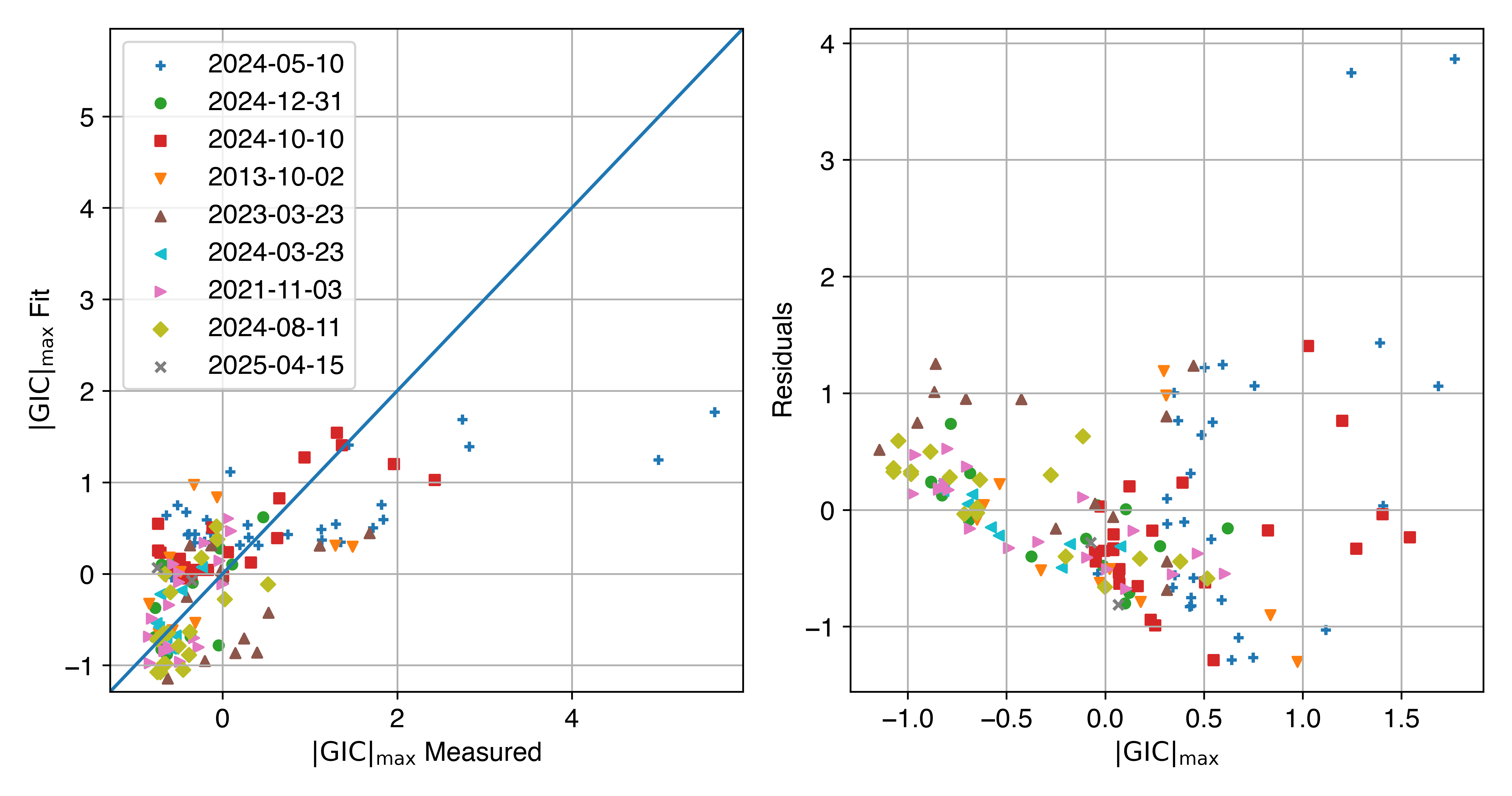}
    \caption{Results from initial regression fit of Equation~\ref{eqn:initialfit} using the nine storms. The plots are in standardized parameters. The left plot shows that the regression under-predicts for large $\gicmax$. Note that the points would lie on the 45-degree blue line if the model perfectly predicted the results.  The right plot shows heteroscedasticity --- the variance in the residuals is smaller at small $\gicmax$. The residuals also have a parabolic shape --- the residuals decrease for $-1 < \gicmax < 0.5$ and they increase for $0.5 < \gicmax < 1.5$. This shape indicates that the response is non-linear.}
    \label{fig:initialfit}
\end{figure}

The results from the initial regression of Equation~\ref{eqn:initialfit} are shown in Figure \ref{fig:initialfit}.  A total of nine storms are included; the storm start dates are shown in the figure legend. The plot of $\gicmax$ from the fit versus the $\gicmax$ measured shows that the regression under-predicts for large $\gicmax$. The residuals plot shows heteroscedasticity. That is, the variance in the residuals is not constant; the residuals have smaller variance at small $\gicmax$.  Common approaches for reducing heteroscedasticity are to apply a log or square root transform to the response variable \cite{james2023statistical}.  In our case, we use $\log_{10}(\gicmax)$. The Breusch-Pagan test (at the 0.05 significance level) shows that this transformation eliminates any statistically significant heteroscedasticity \cite{verma2025comprehensive}. The residuals also have a parabolic shape. That is, the residuals slope downward for $-1 < \gicmax < 0.5$, and they slope upward for $0.5 < \gicmax < 1.5$. This shape indicates that the response is non-linear, and that we should include higher-order terms in the fit \cite{james2023statistical}.  

\subsection{Second-order Fit}
\label{section:secondorder}

Consequently, we fit a second-degree polynomial with all constant, linear, quadratic, and interaction terms:

\begin{equation}
    \begin{split}
        \log_{10}(\gicmax)
        &= A_0 + A_1 \alpha + A_2 \beta + A_3 \bzimfmin + A_4 \vxmax \\
        & + A_5 \alpha^2+ A_6 \beta^2 + A_7 \bzimfmin^2 + A_8 \vxmax^2 \\
        &+ A_9 \alpha \cdot \beta + A_{10} \bzimfmin \cdot \vxmax \\
        &+ A_{11} \alpha \cdot \bzimfmin  + A_{12} \beta \cdot \bzimfmin  \\
        &+ A_{13} \alpha \cdot \vxmax  + A_{14} \beta \cdot \vxmax  + \epsilon
    \end{split}
    \label{eqn:full}
\end{equation}

To avoid overfitting the data, we use best subset selection \cite{james2023statistical}.  We generate a list of all possible combinations of parameters in Equation~\ref{eqn:full}. This will create 15 one-parameter equations, 105 two-parameter equations, \ldots, and one 15-parameter equation.  We create a fit for each equation and determine which equation provides the highest coefficient of determination ($R^2$) for one-parameter equations, the highest $R^2$ for two-parameter equations, etc. Then we examine which of these fits minimizes the corrected Akaike Information Criterion ($\aicc$) \cite{burnham2004multimodel}. 

\begin{figure}
    \centering
    \includegraphics[width=1.0\linewidth]{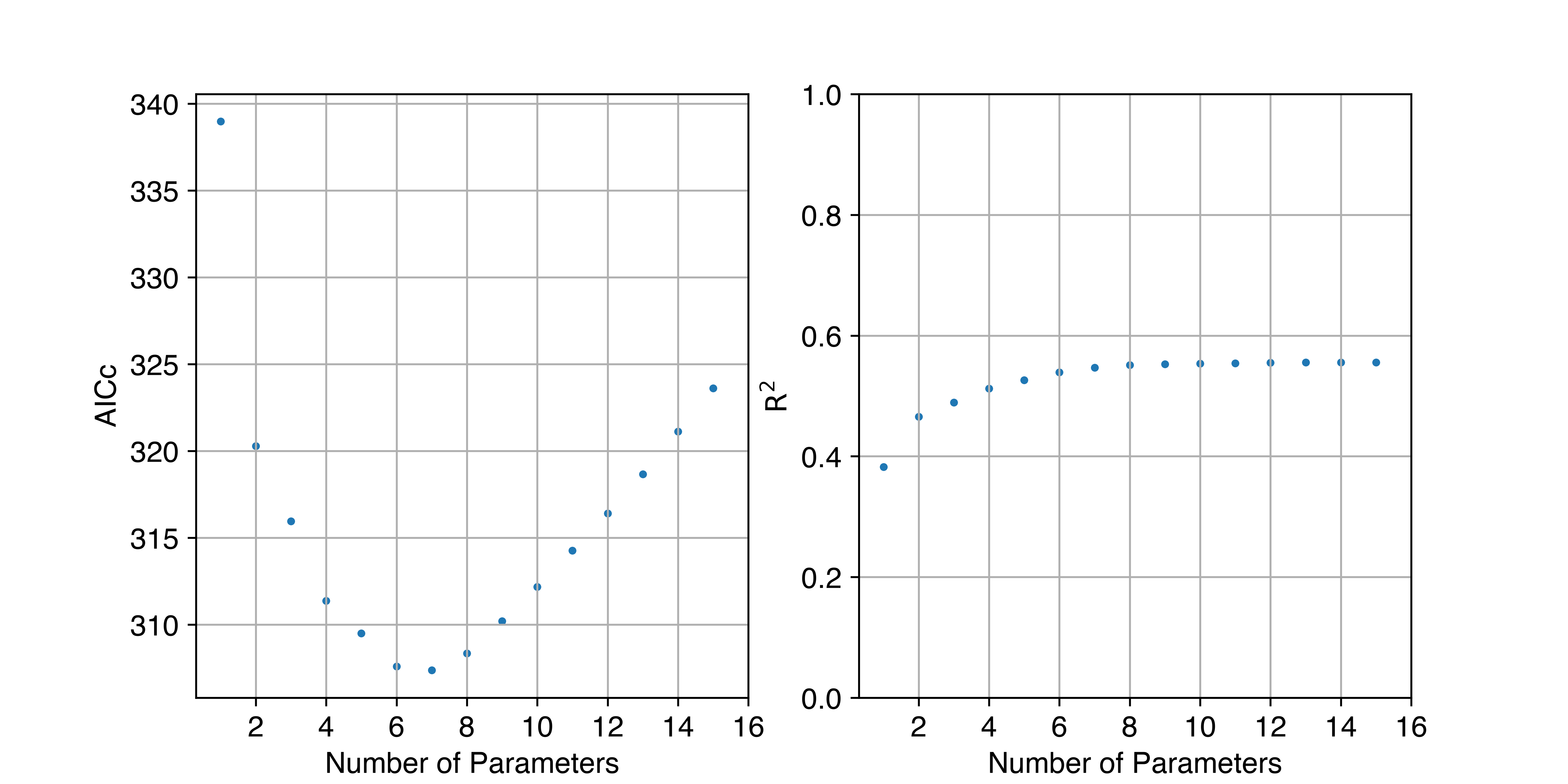}
    \caption{Results from best subset selection for the linear regression of Equation~\ref{eqn:full}.  The left plot shows that the $\aicc$ varies with the number of parameters in the fit. The right plot shows how $R^2$ varies with the number of parameters. The preferred fits are in the $\aicc$ valley between 6 and 8-parameters.}
    \label{fig:aicmetrics}
\end{figure}

Figure~\ref{fig:aicmetrics} shows the results from the best subset selection.  In principle, we select the fit with the lowest $\aicc$, which in this case is the 7-parameter equation, see the left-hand plot in Figure~\ref{fig:aicmetrics}.  However, we note that the fits between 6 and 8-parameters have similar $\aicc$ values.  $\aicc$ values are not interpretable in absolute terms; however, their differences are meaningful. We define $\Delta \aicc_i \equiv \aicc_i - \aicc_{\text{min}}$ where $\aicc_i$ is for the equation with $i$ parameters and $\aicc_{\text{min}}$ is the minimum $\aicc$.   For fits with $\Delta \aicc_i < 2$, the fits have substantial support.  Fits with $\Delta \aicc_i > 10$ have no support \cite{burnham2004multimodel}.  Thus, we cannot easily distinguish between the 6- to 8-parameter fits, which have similar $R^2$s, see the right-hand side of Figure~\ref{fig:aicmetrics}, and explain approximately the same amount of variance in the data.  The ``unstandardized" 6- to 8-parameter equations are:

\begin{equation}
    \begin{split}
    \log_{10}(\gicmax) &= 2.138 + 5.660 \cdot 10^{-2}\bzimfmin + 1.177 \cdot 10^{-3}\vxmax \\
    &- 2.470\alpha + 1.011 \cdot 10^{-3}\bzimfmin^2 - 2.896 \cdot 10^{-3}\alpha \cdot \vxmax \\
    &+ 0.6359\alpha \cdot \beta
    \end{split}
    \label{eqn:6parameter}
\end{equation}

\begin{equation}
    \begin{split}
    \log_{10}(\gicmax) &= 2.073 + 4.833 \cdot 10^{-2}\bzimfmin + 1.238 \cdot 10^{-3}\vxmax \\
    &- 3.982\alpha + 0.2288\beta + 2.997\alpha^2 \\
    &+ 8.965 \cdot 10^{-4}\bzimfmin^2 - 3.311 \cdot 10^{-3}\alpha \cdot \vxmax
    \end{split}
    \label{eqn:7parameter}
\end{equation}

\begin{equation}
    \begin{split}
    \log_{10}(\gicmax) &= 2.228e + 5.404 \cdot 10^{-2}\bzimfmin + 1.148 \cdot 10^{-3}\vxmax \\
    &- 3.553\alpha + 2.851\alpha^2 + 4.680 \cdot 10^{-2}\beta^2 \\
    &+ 9.369 \cdot 10^{-4}\bzimfmin^2 - 1.983 \cdot 10^{-3}\beta \cdot \bzimfmin \\
    &- 2.911 \cdot 10^{-3}\alpha \cdot \vxmax
    \end{split}
    \label{eqn:8parameter}
\end{equation}

These equations are the ``operational" versions, useful to anyone wanting to estimate GICs. The transformation back to unstandardized variables adds a constant term to each equation. Therefore, the 6-parameter equation actually has seven terms. Due to the different scales and units of the unstandardized parameters, the regression coefficients in these equations vary substantially from term to term. In these equations, $\gicmax$ is measured in A, $\bzimfmin$ in nT, and $\vxmax$ in km/s. $\alpha$ and $\beta$ are dimensionless. 

\begin{figure}
    \centering
    \includegraphics[width=1.0\linewidth]{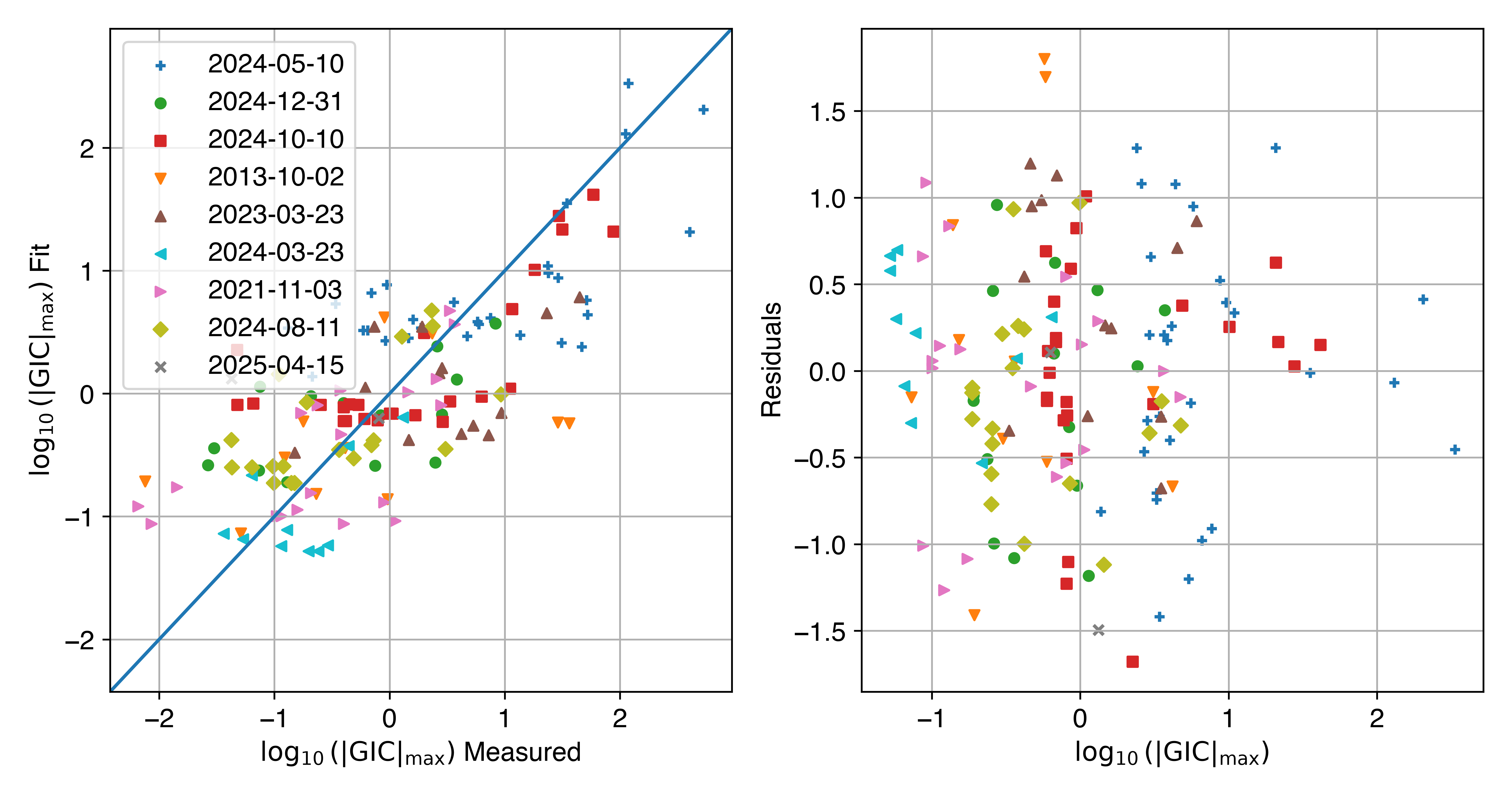}
    \caption{Results for the 6-parameter regression fit, Equation~\ref{eqn:6parameter}.  These plots are in standardized parameters. The left plot shows that the regression fit no longer under-predicts. The data points would lie on the 45-degree blue line if the model perfectly predicted the results. The right plot shows heteroscedasticity is no longer an issue, as confirmed by the Breusch-Pagan test at the 0.05 significance level.}
    \label{fig:aicfit}
\end{figure}

While there is no compelling statistical reason to pick one equation over another, we select the 6-parameter equation because it is parsimonious (an equation with fewer parameters) and has an $\alpha\beta$ term consistent with our earlier analysis that showed a similar dependence. The fit and residuals for this equation are shown in Figure~\ref{fig:aicfit}.  In contrast to the initial fit, this fit no longer under-predicts at large $\gicmax$, and the residuals are homoscedastic (the residuals have constant variance) as confirmed by the Breusch-Pagan test at the 0.05 significance level.

\subsection{Single Storm Influence}
 
We verified that no single storm drives the regression results. Most of the largest $\gicmax$ values are from the Gannon storm, which may influence the regression. To examine this possibility, we performed a ``leave-one-out" cross-validation approach, dropping one storm at a time from the analysis for the 6-parameter equation. This leads to nine regressions with eight storms in each. We determine $R^2$ with all nine storms, $R_i^2$ without storm $i$, and $\Delta R_i^2 = R^2 - R_i^2$, see Table~\ref{tab:cooksdistance}. 

We also determine the sum of Cook's distances ($D_k$) for all points in the dropped storm $i$.  $D_k$ measures influence -- the impact each data point has on the overall regression model -- and is defined as
\begin{equation}
    D_k = \frac{\Sigma_{j=1}^{n} (y_j - y_{j(k)})^2}{p \cdot \text{MSE}},
\end{equation}

\noindent where $y_j$ is the prediction for observation $j$ using the the regression model with all data points, $y_{j(k)}$ is the prediction for observation $j$ from a refitted regression model in which observation $k$  from storm $i$ is dropped, $p$ is the number of parameters in the regression model, and $\text{MSE}$ is the mean squared error of the full regression model.  A $D_k > 1$ indicates that the associated data point is influential \cite{montgomery2019applied}.  The sum of $D_k$ over all points in dropped storm $i$, $\Sigma D_{k \in i}$, estimates each storm's influence over the regression.  

With all $\Sigma D_{k \in i}$ well below one, we demonstrate that the model is not disproportionately influenced by any individual extreme event, such as the Gannon storm.  This is further supported by the fact that we do not see a substantial change in $R^2$.  We conclude that no single storm is driving the regression results.

\begin{table}[]
    \caption{Dropping one storm at a time, we perform nine regressions with eight storms in each.  We calculate $R_i^2$ without the storm $i$, $\Delta R_i^2 = R^2 - R_i^2$, and the total Cook's distance, $\Sigma D_{k \in i}$, for each storm. $\Sigma D_{k \in i}$ measures each storm's influence on the regression.  All $\Sigma D_{k \in i}$s are below one, indicating that none of the storms have excessive influence.  There are a total of 143 observations, divided among the storms as shown in the last column.}
    \centering
\begin{tabular}{ c | c | c |c | c }
Dropped Storm & $R_i^2$ & $\Delta R_i^2$ &  $\Sigma D_{k \in i}$ & Num. Obs.\\
\hline
2024-05-10& 0.474 & 0.066 & 0.268 & 29 \\
2024-03-23& 0.527 & 0.012 & 0.030 & 10 \\
2024-10-10& 0.533 & 0.006 & 0.083 & 25 \\
2021-11-03& 0.534 & 0.005 & 0.106 & 19 \\
2024-12-31& 0.548 & -0.009 & 0.052 & 15 \\
2025-04-15& 0.549 & -0.010 & 0.010 & 2 \\
2024-08-11& 0.550 & -0.011 & 0.178 & 19 \\
2013-10-02& 0.570 & -0.030 & 0.136 & 11 \\
2023-03-23& 0.571 & -0.032 & 0.108 & 13 \\
\end{tabular}
    \label{tab:cooksdistance}
\end{table}

\subsection{Upper Bound}
\label{section:upper}

One of the limitations of the regressions in section~\ref{section:secondorder} is that they estimate the mean $\gicmax$.  During storms, operators are concerned with peak ``worst-case'' values \cite{nerc2025reliability}; so we redo the analysis, performing a quantile regression to identify the upper 80\% bound for $\gicmax$.  This will provide an upper limit to the maximum expected GIC.

To estimate the upper bound, we use a quantile regression rather than the prediction interval for the linear regression.  A prediction interval assumes that the residuals have a normal distribution with constant variance and is sensitive to outliers \cite{koenker2005quantile}. Quantile regression, however, avoids these issues \cite{koenker2005quantile}.  Quantile regression is non-parametric; it does not require any assumption about the distribution of the residuals. The residuals do not need to be normally distributed and can be skewed. Additionally, quantile regression models are generally less sensitive to outliers than linear regression models.  Because it has fewer limitations, we employ quantile regression to estimate the upper bound.

We follow the same procedure as Section~\ref{section:secondorder} with a few changes.  We use the statsmodels quantile regression function in a best subset selection. AICc is not defined for quantile regression, so we use a pseudo-$\aicc$ \cite{cade2005quantile} defined as:
\begin{equation}
    \pseudoaicc = n \cdot \log(C_f / n) + 2k + 2(k^2 + k)/(n-k-1)
\end{equation}
\noindent
where $n$ is the number of observations, $k$ is the number of parameters in the equation, and $C_f$ is the check function, also known as pinball loss.  This definition of pseudo-${\aicc}$ includes a correction for small sample sizes.  We use pseudo-${\aicc}$ and pseudo $R^2$ (as reported by statsmodels) for the best subset selection.

\begin{figure}
    \centering
    \includegraphics[width=1.0\linewidth]{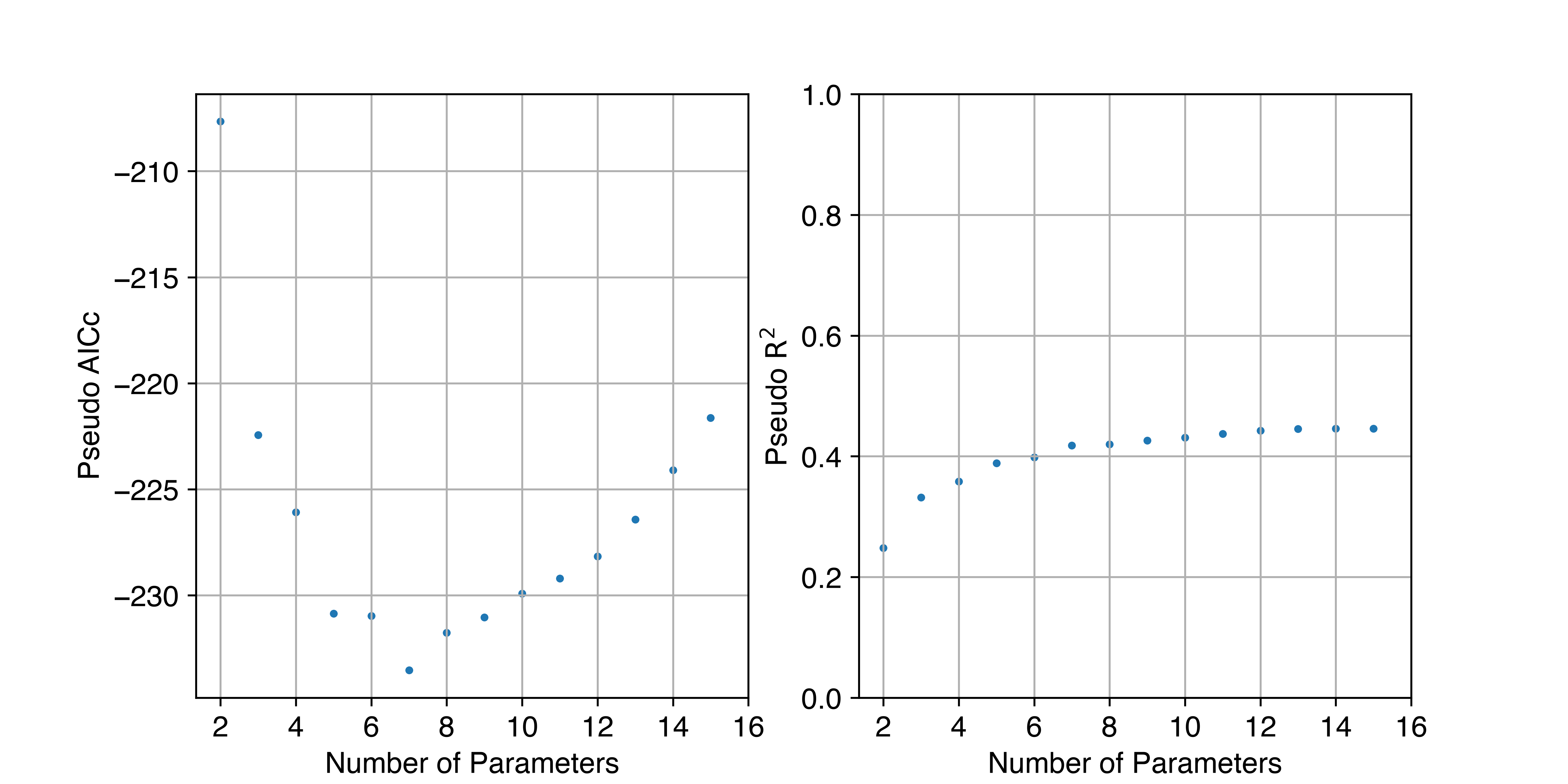}
    \caption{Results from best subset selection for the quantile regression.  The left plot shows that the $\pseudoaicc$ varies with the number of parameters in the fit. The right plot shows how pseudo $R^2$ varies with the number of parameters. The preferred fits are in the $\pseudoaicc$ valley between 5 and 9-parameters. }
    \label{fig:quantmetrics}
\end{figure}

\begin{figure}
    \centering
    \includegraphics[width=1.0\linewidth]{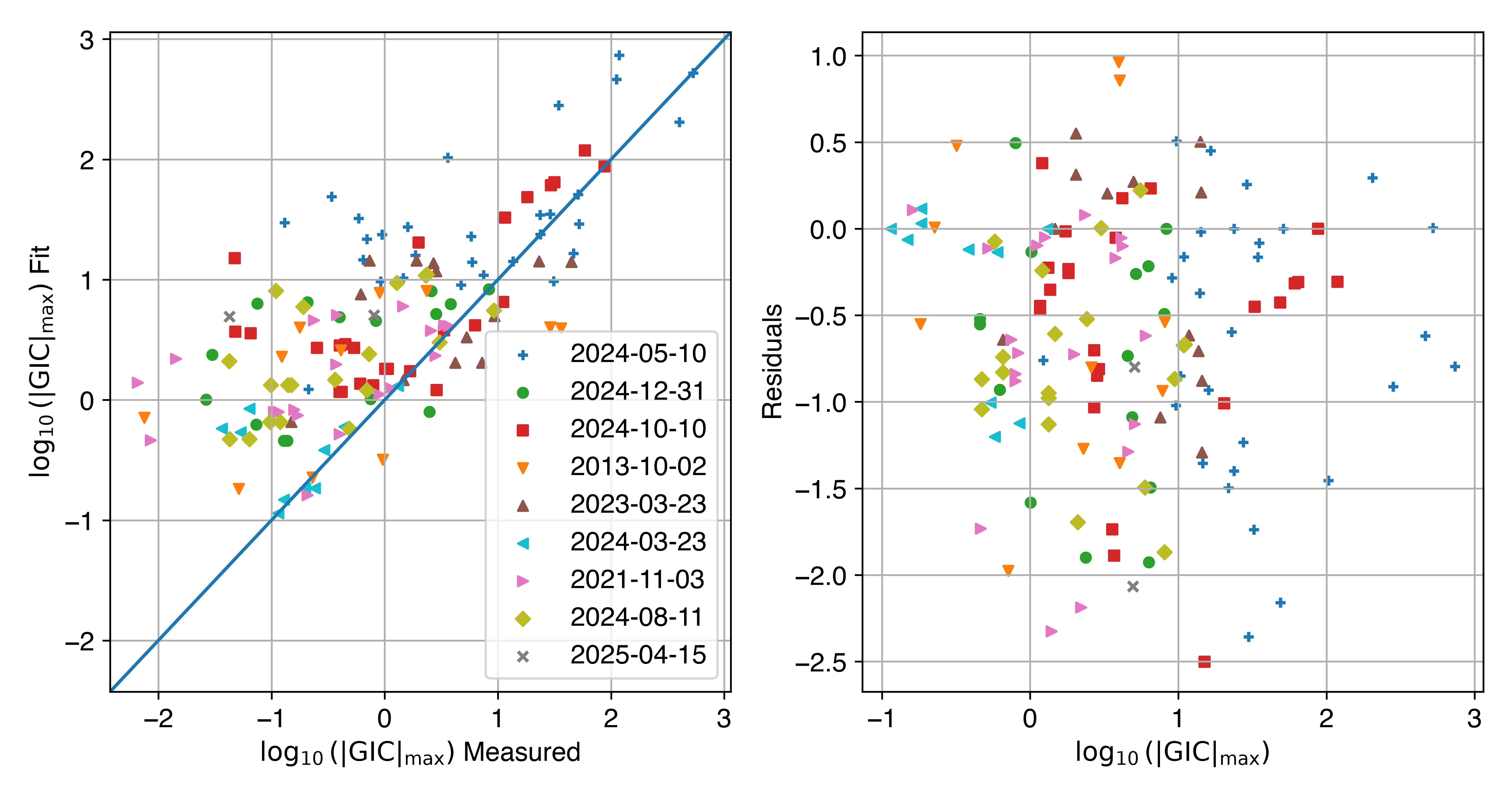}
    \caption{Results from the quantile best subset selection, Equation~\ref{eqn:upper80}. The plots are in standardized parameters.  The points are mostly above the diagonal line because this fit is the upper 80\% bound.}
    \label{fig:quantfit}
\end{figure}

Equation~\ref{eqn:upper80} is the unstandardized upper 80\% bound for $\gicmax$.  This 7-parameter equation is at the bottom of the pseudo-${\aicc}$ valley between 5 and 9-parameters, see Figure~\ref{fig:quantmetrics}. The fit for the standardized equation is shown in Figure~\ref{fig:quantfit}.  On the left side, the points are mostly above the diagonal line because this equation represents an upper bound. 
\begin{equation}
\begin{split}
\log_{10}(\gicmaxub) &= 1.597 + 7.605 \cdot 10^{-2} \bzimfmin + 0.5114 \beta \\
&- 8.607 \cdot 10^{-2} \beta^2 + 1.121 \cdot 10^{-3} \bzimfmin^2 \\
&- 4.700 \cdot 10^{-2} \alpha \cdot \bzimfmin + 1.728 \cdot 10^{-3} \alpha \cdot \vxmax
\end{split}
\label{eqn:upper80}
\end{equation}
\noindent Note that the 7-parameter equation already had a constant term, so unstandardizing the parameters did not change the number of terms in the equation.

A reasonable question is why use the upper 80\% bound rather than, for example, the upper 95\% bound. The same process could be used to determine an upper 95\% bound; however, we consider the estimated upper 95\% bound to be unrealistic. The equation for the upper 95\% bound predicts $\gicmax$ values exceeding 10,000 [A], far in excess of any values in the data set, see \ref{app:data}. This unrealistic result is due to sparse data at the upper 95\% bound. With 143 observations in our data set, $0.05 \times 143 \approx 7$ data points drive the estimate of the upper 95\% bound. As \citeA{koenker2005quantile} notes, when the data are sparse at the quantile of interest, it is difficult to precisely estimate the quantile. In contrast, the upper 80\% bound is well-supported by the data, with $0.2 \times 143 \approx 29$ data points supporting the fit. It provides reasonable $\gicmax$ estimates consistent with values in the data set. Consequently, we limit the analysis to the upper 80\% bound.

\subsection{Storm statistics}
\label{section:statistics}

We want to understand the $\gicmax$ statistical distribution for the upper bounds. This information is useful, for example, in creating Monte Carlo simulations looking at the impact of solar storms on technological infrastructure or using Extreme Value Theory to estimate 1-in-100 or 1-in-200-year $\gicmax$ values. Following the process outlined below, we use Equation~\ref{eqn:upper80} to examine the $\gicmax$ distribution, which we will show is consistent with a lognormal distribution. 

Two of the four parameters in Equation~\ref{eqn:upper80} are $\alpha$ and $\beta$. To determine a range of $\alpha$ and $\beta$ values, we examined the nine-storm analysis of sections~\ref{section:initial} through \ref{section:upper}. That analysis has 67 unique sites from which we extract 67 pairs of $\alpha$ and $\beta$ values.

The other two parameters are $\bzimfmin$ and $\vxmax$.  To determine a range of $\bzimfmin$ and $\vxmax$ values, we identified all $K_p \ge 8$ storms in OMNI data since 1995, \cite{matzka2021geomagnetic, OMNI2020_1min}. To capture leading and trailing edges of the storms, we include $K_p \ge 6$ periods before or after each $K_p \ge 8$ period. This process identified 39 storms since 1995. For each storm, we determine $\bzimfmin$ and $\vxmax$. 

\begin{figure}[h]
    \centering
    \includegraphics[width=0.65\linewidth]{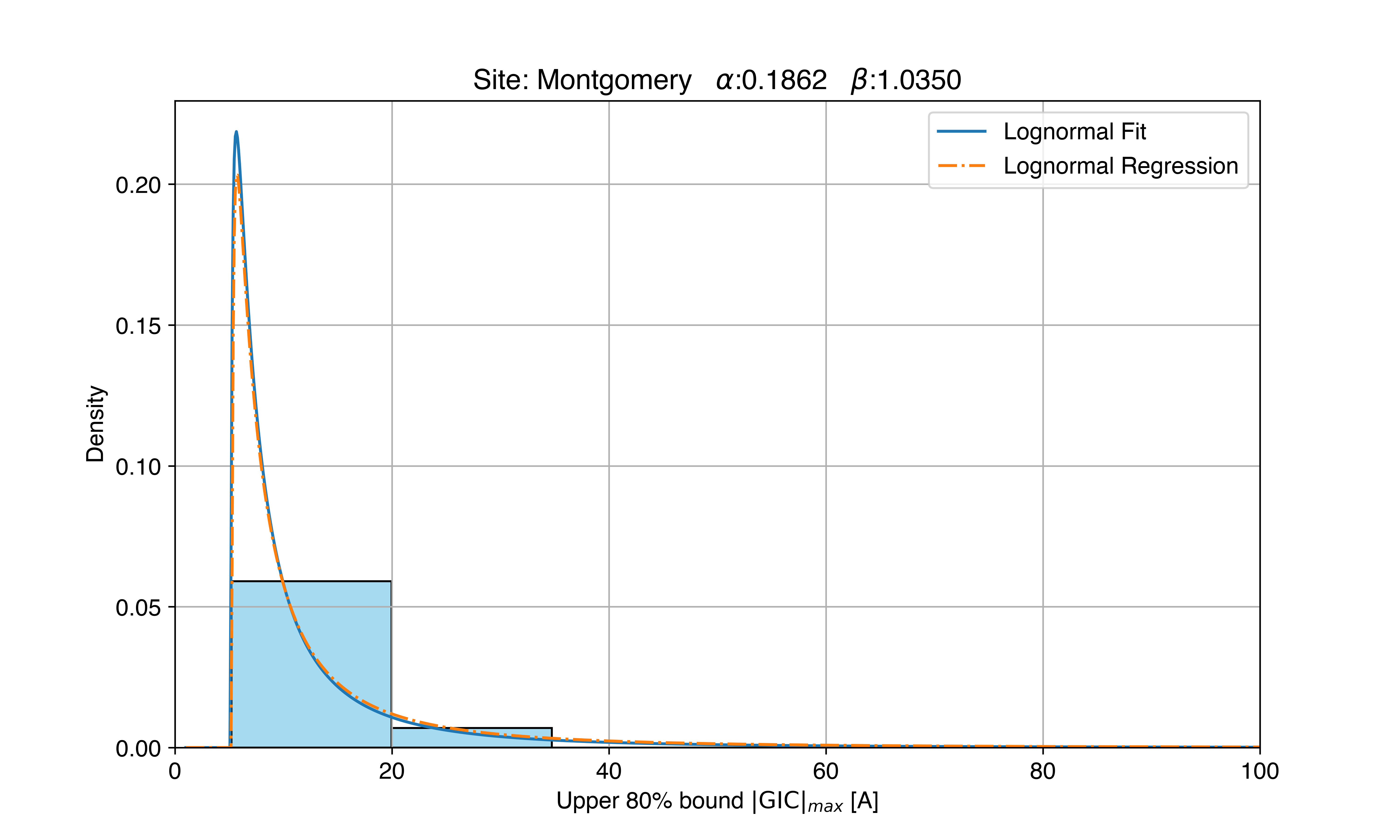}
    \caption{Distribution of upper 80\% bounds for $\gicmax$ for Montgomery based on Equation~\ref{eqn:upper80} and 39 $K_p \ge 8$ storms since 1995. This plot has smaller $\gicmax$ values than Figure~\ref{fig:10675fit} primarily because the site is further south (smaller $\alpha$). The blue line is a lognormal fitted to the data; the dashed red line is a lognormal distribution based on Equations~\ref{eqn:lognormal_parameters}. The two lines nearly overlap.}
    \label{fig:Montgomeryfit}
\end{figure}

\begin{figure}[h]
    \centering
    \includegraphics[width=0.65\linewidth]{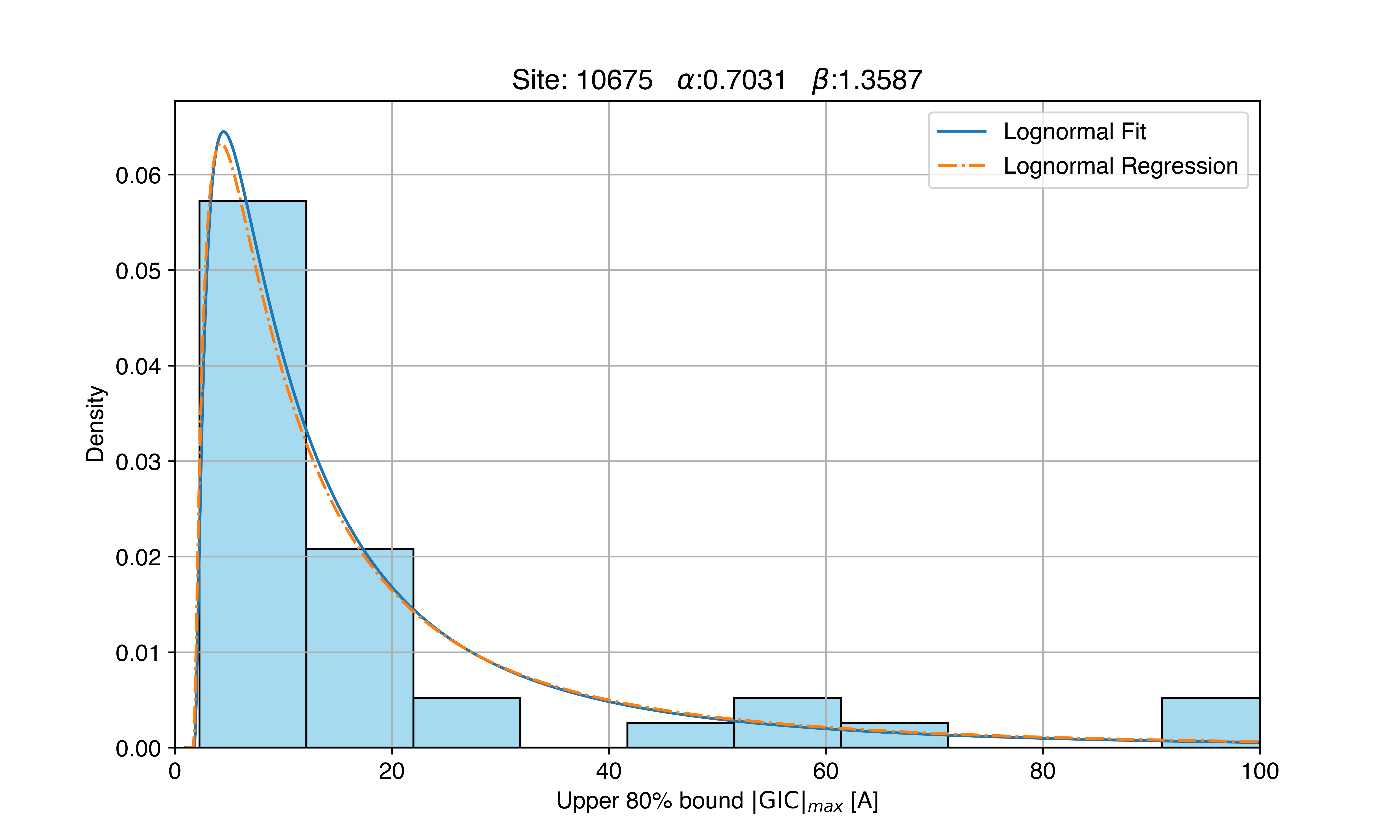}
    \caption{Distribution of upper 80\% bounds for $\gicmax$ for site 10675 based on Equation~\ref{eqn:upper80} and 39 $K_p \ge 8$ storms  since 1995. The distribution has larger $\gicmax$ values than Figure~\ref{fig:Montgomeryfit} primarily because the site is further north (larger $\alpha$). Otherwise, the description is the same as Figure~\ref{fig:Montgomeryfit}.}
    \label{fig:10675fit}
\end{figure}

Using Equation~\ref{eqn:upper80} and the above values, we determine the upper 80\% bound for $\gicmax$ for all 39 storms for each site and plot the results in a histogram. With 67 sites, we get 67 histograms.  Figures~\ref{fig:Montgomeryfit} and \ref{fig:10675fit} contain the histograms for two sites. There is nothing special about these sites, and the results are similar for all sites. 

We considered lognormal, gamma, and Weibull distributions as possible fits to the data. For all sites, the Kolmogorov-Smirnov (K-S) test \cite{heckert2002handbook} indicates the lognormal distribution has acceptable goodness-of-fit to the data, tested at the 0.05 significance level. The gamma and Weibull distributions were rejected at many sites. Consequently, we use a lognormal distribution to describe the distributions, and using the scipy lognormal function \cite{2020SciPy-NMeth}, we create lognormal fits to the data in each histogram. Figures~\ref{fig:Montgomeryfit} and \ref{fig:10675fit} show the lognormal fits for those sites. 

The lognormal fits can be generalized for an arbitrary site with a known $\alpha$ and $\beta$. We created a 20 by 20 grid of $\alpha$ and $\beta$ parameters uniformly covering the range of $\alpha$s and $\beta$s seen in the nine-storm data. 

For each of these 400 $\alpha \beta$ pairs, we created a histogram similar to those in Figures~\ref{fig:Montgomeryfit} and \ref{fig:10675fit}.  We fit the scipy lognormal to each of the histograms.  The scipy lognormal probability density function has three parameters -- shape ($\sigma$), location ($\theta$), and scale ($m$) -- and is represented by:

\begin{equation}
    f(\gicmax,\sigma,\theta,m) = \frac{e^{-\frac{\big[\log \big(\frac{\tiny{\gicmax}-\theta}{m} \big)\big]^2}{2 \sigma^2}}}{\sqrt{2 \pi}  (\gicmax-\theta) \sigma} 
\end{equation}

\begin{figure}[t]
    \centering
    \includegraphics[width=1.0\linewidth]{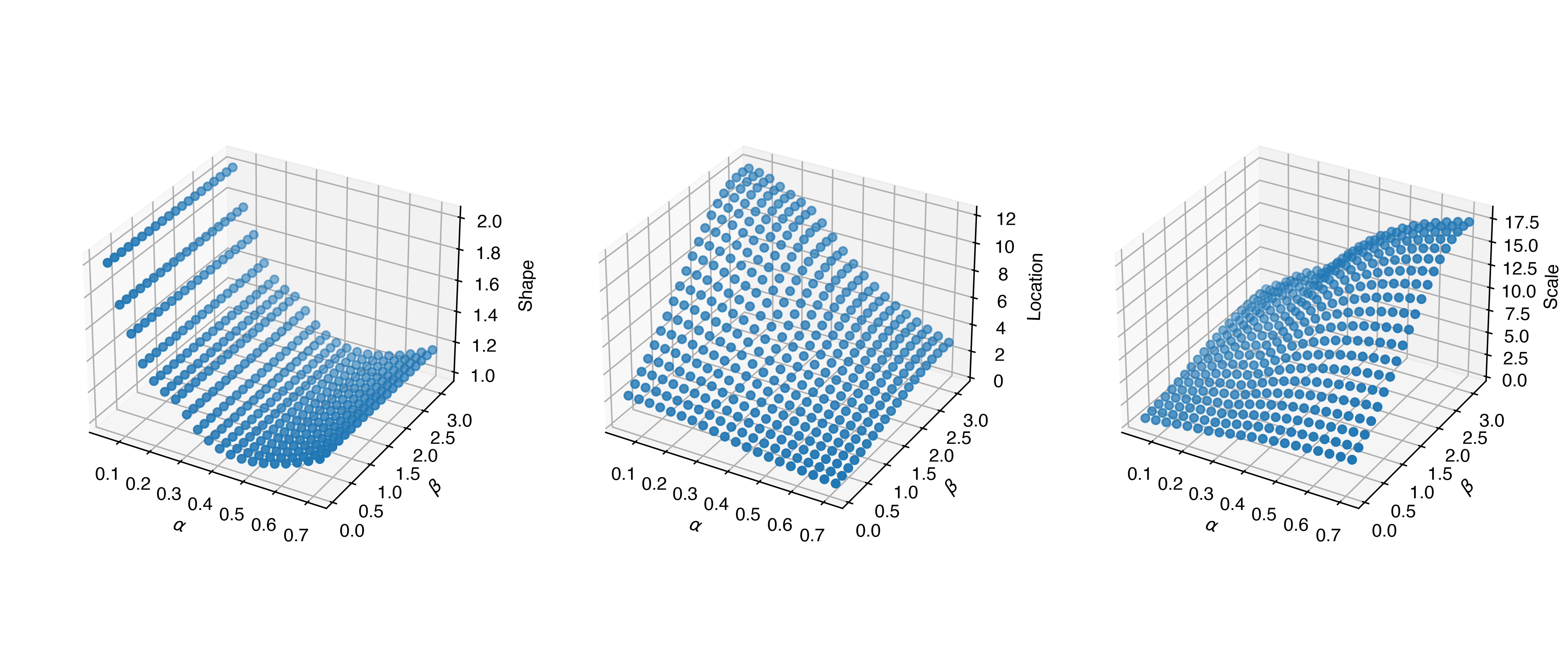}
    \caption{Shape, location, and scale parameters for lognormal distributions as a function of $\alpha$ and $\beta$. Equations~\ref{eqn:lognormal_parameters} are regression fits to these surfaces, and these equations can be used to estimate shape, location, and scale parameters for the lognormal distribution. The resultant lognormal distribution is a good fit to the distribution of upper 80\% bounds for $\gicmax$.}
    \label{fig:params}
\end{figure}

For all 400 histograms, we determined the shape, location, and scale parameters for the associated lognormal distribution, which are shown in Figure~\ref{fig:params}. We note that the parameters change smoothly across the space, and we use best subset selection to find regression fits:
\begin{equation}
\begin{split}
    \sigma &=  2.100 -4.525 \alpha + 4.617 \alpha^2 \\
    \theta &=  1.908 - 3.331 \alpha + 4.937 \beta + 0.743 \alpha^2 -0.314 \beta^2 -4.580 \alpha \beta \\
    m &=  -1.096 + 9.785 \alpha + 2.478 \beta -7.615 \alpha^2 -0.491 \beta^2 + 6.620 \alpha \beta
\end{split}
\label{eqn:lognormal_parameters}
\end{equation}
These equations capture almost all the variance in the shape, location, and scale parameters.  The three equations have a Pearson correlation coefficient, $r$, of 0.99 and an $R^2=0.98$.

Equations~\ref{eqn:lognormal_parameters} can be used to determine the shape, location, and scale parameters for an arbitrary site with known $\alpha$ and $\beta$. Figures~\ref{fig:Montgomeryfit} and \ref{fig:10675fit} contain two lines for the lognormal distribution.  One line is the lognormal fit to the histogram in the figure.  The other line uses Equations~\ref{eqn:lognormal_parameters} to estimate the parameters for the lognormal distribution. The two lines almost overlap, showing that Equations~\ref{eqn:lognormal_parameters} provide a good representation of the lognormal statistics at an arbitrary site. 
 

\newpage

\section{Conclusions}

This work provides fast, computationally efficient methods for estimating the mean $\gicmax$ in a solar storm, the upper 80\% bound on $\gicmax$, and the distribution of those upper bounds. Equations:
\begin{equation}\tag{5}
    \begin{split}
    \log_{10}(\gicmax) &= 2.138 + 5.660 \cdot 10^{-2}\bzimfmin + 1.177 \cdot 10^{-3}\vxmax \\
    &- 2.470\alpha + 1.011 \cdot 10^{-3}\bzimfmin^2 - 2.896 \cdot 10^{-3}\alpha \cdot \vxmax \\
    &+ 0.6359\alpha \cdot \beta
    \end{split}
\end{equation}
\begin{equation}\tag{10}
\begin{split}
\log_{10}(\gicmaxub) &= 1.597 + 7.605 \cdot 10^{-2} \bzimfmin + 0.5114 \beta \\
&- 8.607 \cdot 10^{-2} \beta^2 + 1.121 \cdot 10^{-3} \bzimfmin^2 \\
&- 4.700 \cdot 10^{-2} \alpha \cdot \bzimfmin + 1.728 \cdot 10^{-3} \alpha \cdot \vxmax
\end{split}
\end{equation}
\noindent allow the user to estimate \noindent mean $\gicmax$ and the upper 80\% bound for $\gicmax$. And following the approach of Section~\ref{section:statistics}, a user can estimate the distribution of the upper 80\% bounds for a given site over $K_p \ge 8$ solar wind conditions observed since 1995.  

The equations are based on data from nine solar storms. Equation~\ref{eqn:6parameter} has an $r=0.73$ and an $R^2$ (uncentered) of 0.54.  Our equations account for storm intensity through $\bzimfmin$ and $\vxmax$, which are associated with geomagnetic activity and geomagnetic storms \cite{burton1975empirical, santoso2024intensity}.  Consistent with our results, \citeA{figueroa2025geomagnetic} have shown that storm intensity and average magnetic field strength are statistically significant predictors of power outages. Our equations also account for local conditions through $\alpha$ and $\beta$. The $\alpha$ scaling factor correlates the peak geoelectric field to geomagnetic latitude, while the $\beta$ scaling factor addresses how the ground's conductivity varies across Earth's surface \cite{nerc2025reliability}. 

We note that Equation~\ref{eqn:6parameter} for mean $\gicmax$ is a multivariate equation that contains a term proportional to $\alpha \beta$.  This is consistent with the recently published paper \cite{wilkerson} that found a regression based solely on the Gannon storm data resulted in an equation with an $\alpha\beta$ term. That earlier analysis, which also examined the mean $\gicmax$, did not include the strength of the storm and consequently did not include the terms involving $\bzimfmin$ and $\vxmax$. 

Unlike existing frameworks that focus on the geoelectric field and detailed knowledge of the power grid (e.g., lengths of power lines, power line interconnections), our methodology provides an operational, site-specific GIC estimation tool. By bridging the gap between solar wind parameters and standardized NERC scaling factors, this empirical approach offers a computationally efficient alternative to physics-based simulations, which is suitable for real-time risk assessment. The estimates from our methodology are appropriate for high-level analysis where the power grid configuration may not be known.

Along these lines, our formulation differs from other analyses; we provide a means of estimating GICs rather than describing the characteristics of the geoelectric field or more general metrics such as $Dst$.  As examples, others have examined the statistical relationship between the solar wind and metrics measured on Earth. \citeA{si2022impacts} examined GMDs and observed that the induced geoelectric field has a lognormal distribution.  \citeA{pulkkinen2008statistics} described the statistics of the geoelectric field conditioned by the state of the magnetosphere and the solar wind.  In addition, \citeA{santoso2024intensity} statistically examined the minimum disturbance storm time index ($Dst_{\text{min}}$) and solar wind conditions. While we provide additional tools, our results are consistent with these analyses.

\citeA{nerc2025reliability} contains a section on Transmission System Planned Performance for Geomagnetic Disturbance Events.  That section highlights the risks associated with these events, noting that GICs, ``may cause transformer hot-spot heating or damage, loss of Reactive Power sources, increased Reactive Power demand, and Misoperation(s)\ldots"  The standard specifies that ``regional geoelectric field peak amplitude used in GMD Vulnerability Assessment, $E_{peak}$, can be obtained from the reference geoelectric field value of 8 V/km for the benchmark GMD event (1) or 12 V/km for the supplemental GMD event."  This process describes the electric field and not the GIC.  Determining GIC would require additional analysis utilizing the power system configuration as shown by \citeA{boteler2014methodology}.  While the NERC equations examine only the peak electric field, our analysis is consistent with their work.

We do not include a factor for power system configuration in our analysis, so it's not surprising that we do not capture all the variance in the GIC data.  Adding information on the power system configuration might improve the fit.  

%
%

\section*{Open Research Section}
The data used in this paper are provided in \ref{app:data}. The Python code used in the analysis is available at \cite{Thomas_Github}.

\section*{Conflict of Interest declaration}
The authors declare there are no conflicts of interest for this manuscript.

\acknowledgments
We acknowledge TVA and NERC ERO Portal (NERC, 2024) and its data providers for furnishing data for this analysis. This includes contributions from the Electric Power Research Institute (EPRI) SUNBURST network. This work is supported by funding from the National Science Foundation (NSF) RAPID Grant (2434136). We also acknowledge the United States Geological Survey (USGS) and Natural Resources Canada (NRCan) for funding, operating, maintaining, and providing the operational magnetometer data used in this study. We also thank the funding agencies and investigators who have openly provided their magnetotelluric transfer function products, specifically USArray (NSF) and USMTArray (USGS and National Air and Space Administration (NASA)).

\appendix
\section{GIC, Solar Wind, and Scaling Factors} 
\label{app:data}

Table~\ref{tab:data} contains the data from the nine storms used in the analysis.  The table contains both NERC and TVA data. Nine of the sites are at the same location, although they have different NERC and TVA Site IDs.  NERC Site IDs are numbers; TVA Site IDs are names.  The matching Site IDs are 10197 and Sullivan, 10204 and Shelby, 10208 and Rutherford, 10203 and Raccoon Mountain, 10212 and Pinhook, 10201 and Montgomery, 10660 and Gleason, 10200 and East Point, and 10207 and Bull Run. 

\begin{longtblr}[
  caption = {Data from the nine storms used in the linear regressions},
  label = {tab:data},]
    {
      colspec = {*{7}{Q[c,m]}},
      rowhead = 1,
      row{1} = {gray9},
    } 
$\gicmax$ [A]& $\bzimfmin$ [nT]& $\vxmax$ [km/s]& $\alpha$ & $\beta$ & Site ID & Date \\
45.4529 & -55.8100 & -1027.1000 & 0.3364 & 2.7748 & 10099 & 2024-05-10 \\
46.4000 & -55.8100 & -1027.1000 & 0.5029 & 2.0699 & 10255 & 2024-05-10 \\
27.5883 & -55.8100 & -1027.1000 & 0.3264 & 0.7867 & 10079 & 2024-05-10 \\
32.8500 & -55.8100 & -1027.1000 & 0.1073 & 0.9648 & 10107 & 2024-05-10 \\
25.6167 & -55.8100 & -1027.1000 & 0.4982 & 0.2891 & 10115 & 2024-05-10 \\
34.0417 & -55.8100 & -1027.1000 & 0.1954 & 1.3052 & 10428 & 2024-05-10 \\
25.5508 & -55.8100 & -1027.1000 & 0.4133 & 0.5914 & 10438 & 2024-05-10 \\
81.0033 & -55.8100 & -1027.1000 & 0.3563 & 2.8969 & 10181 & 2024-05-10 \\
4.4517 & -55.8100 & -1027.1000 & 0.0547 & 0.0771 & 10182 & 2024-05-10 \\
73.0967 & -55.8100 & -1027.1000 & 0.2538 & 2.0985 & 10184 & 2024-05-10 \\
16.6917 & -55.8100 & -1027.1000 & 0.2606 & 0.4584 & 10185 & 2024-05-10 \\
7.7467 & -55.8100 & -1027.1000 & 0.3568 & 0.5617 & 10186 & 2024-05-10 \\
6.9050 & -55.8100 & -1027.1000 & 0.3165 & 0.6117 & 10187 & 2024-05-10 \\
29.4133 & -55.8100 & -1027.1000 & 0.2764 & 2.3549 & 10195 & 2024-05-10 \\
6.5000 & -55.8100 & -1027.1000 & 0.1309 & 1.2388 & 10249 & 2024-05-10 \\
12.7400 & -55.8100 & -1027.1000 & 0.1435 & 1.9962 & 10250 & 2024-05-10 \\
14.0580 & -55.8100 & -1027.1000 & 0.1769 & 0.5554 & Bull Run & 2024-05-10 \\
3.7163 & -55.8100 & -1027.1000 & 0.1417 & 1.1660 & East Point & 2024-05-10 \\
15.1657 & -55.8100 & -1027.1000 & 0.1775 & 0.9440 & Gleason & 2024-05-10 \\
9.4095 & -55.8100 & -1027.1000 & 0.1738 & 1.0396 & Johnsonville & 2024-05-10 \\
34.3022 & -55.8100 & -1027.1000 & 0.1862 & 1.0350 & Montgomery & 2024-05-10 \\
15.3057 & -55.8100 & -1027.1000 & 0.2012 & 0.6773 & Paradise 3 & 2024-05-10 \\
9.9688 & -55.8100 & -1027.1000 & 0.1754 & 0.7896 & Pinhook & 2024-05-10 \\
6.7168 & -55.8100 & -1027.1000 & 0.1711 & 0.7617 & Rutherford & 2024-05-10 \\
9.0932 & -55.8100 & -1027.1000 & 0.1597 & 0.6429 & Shelby & 2024-05-10 \\
7.6637 & -55.8100 & -1027.1000 & 0.1525 & 0.6305 & Southaven & 2024-05-10 \\
5.2915 & -55.8100 & -1027.1000 & 0.1870 & 1.3051 & Sullivan & 2024-05-10 \\
28.2972 & -55.8100 & -1027.1000 & 0.1431 & 0.6530 & Union & 2024-05-10 \\
20.8235 & -55.8100 & -1027.1000 & 0.1542 & 0.7870 & Widows Creek & 2024-05-10 \\
13.0000 & -23.3500 & -552.5000 & 0.5012 & 2.0699 & 10255 & 2024-12-31 \\
11.0950 & -23.3500 & -552.5000 & 0.1070 & 0.9648 & 10107 & 2024-12-31 \\
5.6220 & -23.3500 & -552.5000 & 0.5717 & 1.7879 & 10676 & 2024-12-31 \\
11.2493 & -23.3500 & -552.5000 & 0.5148 & 2.3574 & 10677 & 2024-12-31 \\
2.0550 & -23.3500 & -552.5000 & 0.1727 & 1.0498 & 10661 & 2024-12-31 \\
3.7917 & -23.3500 & -552.5000 & 0.1993 & 0.6881 & 10691 & 2024-12-31 \\
3.6933 & -23.3500 & -552.5000 & 0.1993 & 0.6881 & 10692 & 2024-12-31 \\
17.3467 & -23.3500 & -552.5000 & 0.3469 & 3.1458 & 10180 & 2024-12-31 \\
11.6483 & -23.3500 & -552.5000 & 0.2529 & 2.0985 & 10184 & 2024-12-31 \\
2.1500 & -23.3500 & -552.5000 & 0.3123 & 1.4474 & 10191 & 2024-12-31 \\
7.4050 & -23.3500 & -552.5000 & 0.3429 & 1.8959 & 10193 & 2024-12-31 \\
3.0233 & -23.3500 & -552.5000 & 0.3847 & 2.1945 & 10194 & 2024-12-31 \\
4.4100 & -23.3500 & -552.5000 & 0.2754 & 2.3549 & 10195 & 2024-12-31 \\
7.1050 & -23.3500 & -552.5000 & 0.1836 & 1.0500 & 10201 & 2024-12-31 \\
3.0017 & -23.3500 & -552.5000 & 0.1522 & 0.8374 & 10209 & 2024-12-31 \\
41.5667 & -47.6900 & -826.0000 & 0.5018 & 2.0699 & 10255 & 2024-10-10 \\
10.2167 & -47.6900 & -826.0000 & 0.4539 & 1.2127 & 10315 & 2024-10-10 \\
9.5700 & -47.6900 & -826.0000 & 0.1783 & 0.7135 & 10728 & 2024-10-10 \\
19.3333 & -47.6900 & -826.0000 & 0.1950 & 1.3052 & 10428 & 2024-10-10 \\
15.6358 & -47.6900 & -826.0000 & 0.4123 & 0.5914 & 10438 & 2024-10-10 \\
35.7953 & -47.6900 & -826.0000 & 0.5155 & 2.3574 & 10677 & 2024-10-10 \\
6.2767 & -47.6900 & -826.0000 & 0.3158 & 0.6117 & 10646 & 2024-10-10 \\
5.6300 & -47.6900 & -826.0000 & 0.1762 & 0.9314 & 10660 & 2024-10-10 \\
2.8767 & -47.6900 & -826.0000 & 0.1729 & 1.0498 & 10661 & 2024-10-10 \\
7.9283 & -47.6900 & -826.0000 & 0.1995 & 0.6881 & 10691 & 2024-10-10 \\
8.0883 & -47.6900 & -826.0000 & 0.1995 & 0.6881 & 10692 & 2024-10-10 \\
5.7183 & -47.6900 & -826.0000 & 0.1761 & 0.5593 & 10693 & 2024-10-10 \\
27.7400 & -47.6900 & -826.0000 & 0.3474 & 3.1458 & 10180 & 2024-10-10 \\
28.4700 & -47.6900 & -826.0000 & 0.3554 & 2.8969 & 10181 & 2024-10-10 \\
6.5683 & -47.6900 & -826.0000 & 0.2600 & 0.4584 & 10185 & 2024-10-10 \\
5.8817 & -47.6900 & -826.0000 & 0.3560 & 0.5617 & 10186 & 2024-10-10 \\
4.7400 & -47.6900 & -826.0000 & 0.3158 & 0.6117 & 10187 & 2024-10-10 \\
2.5517 & -47.6900 & -826.0000 & 0.3127 & 1.4474 & 10191 & 2024-10-10 \\
19.6267 & -47.6900 & -826.0000 & 0.3434 & 1.8959 & 10193 & 2024-10-10 \\
2.5600 & -47.6900 & -826.0000 & 0.1565 & 1.1070 & 10199 & 2024-10-10 \\
12.3850 & -47.6900 & -826.0000 & 0.1838 & 1.0500 & 10201 & 2024-10-10 \\
5.6283 & -47.6900 & -826.0000 & 0.1761 & 0.5593 & 10207 & 2024-10-10 \\
7.2450 & -47.6900 & -826.0000 & 0.1500 & 0.6618 & 10210 & 2024-10-10 \\
11.7117 & -47.6900 & -826.0000 & 0.1415 & 0.6415 & 10211 & 2024-10-10 \\
23.2000 & -47.6900 & -826.0000 & 0.3060 & 2.7192 & 10383 & 2024-10-10 \\
4.1667 & -32.5600 & -632.9000 & 0.4592 & 1.6600 & 10301 & 2013-10-02 \\
3.6333 & -32.5600 & -632.9000 & 0.4805 & 1.2746 & 10318 & 2013-10-02 \\
5.6833 & -32.5600 & -632.9000 & 0.4462 & 1.3986 & 10319 & 2013-10-02 \\
29.9500 & -32.5600 & -632.9000 & 0.2738 & 2.0732 & 10397 & 2013-10-02 \\
4.5867 & -32.5600 & -632.9000 & 0.1202 & 0.6517 & 10331 & 2013-10-02 \\
7.6100 & -32.5600 & -632.9000 & 0.3716 & 3.1458 & 10180 & 2013-10-02 \\
10.8433 & -32.5600 & -632.9000 & 0.3802 & 2.8969 & 10181 & 2013-10-02 \\
2.6267 & -32.5600 & -632.9000 & 0.3457 & 0.3383 & 10183 & 2013-10-02 \\
27.5367 & -32.5600 & -632.9000 & 0.2707 & 2.0985 & 10184 & 2013-10-02 \\
7.7933 & -32.5600 & -632.9000 & 0.2127 & 0.6881 & 10198 & 2013-10-02 \\
1.2900 & -32.5600 & -632.9000 & 0.1669 & 1.1070 & 10199 & 2013-10-02 \\
6.6000 & -18.4700 & -516.0000 & 0.3868 & 1.8850 & 10311 & 2023-03-23 \\
18.1000 & -18.4700 & -516.0000 & 0.3937 & 1.5721 & 10321 & 2023-03-23 \\
14.6833 & -18.4700 & -516.0000 & 0.1967 & 1.3052 & 10428 & 2023-03-23 \\
10.0467 & -18.4700 & -516.0000 & 0.2948 & 2.9734 & 10647 & 2023-03-23 \\
13.4483 & -18.4700 & -516.0000 & 0.1744 & 1.0498 & 10661 & 2023-03-23 \\
32.3633 & -18.4700 & -516.0000 & 0.3504 & 3.1458 & 10180 & 2023-03-23 \\
25.3383 & -18.4700 & -516.0000 & 0.3586 & 2.8969 & 10181 & 2023-03-23 \\
7.0533 & -18.4700 & -516.0000 & 0.2948 & 2.9734 & 10188 & 2023-03-23 \\
11.6583 & -18.4700 & -516.0000 & 0.2782 & 2.3549 & 10195 & 2023-03-23 \\
16.4817 & -18.4700 & -516.0000 & 0.1854 & 1.0500 & 10201 & 2023-03-23 \\
9.1167 & -18.4700 & -516.0000 & 0.1775 & 0.8957 & 10205 & 2023-03-23 \\
3.9100 & -18.4700 & -516.0000 & 0.1776 & 0.5593 & 10207 & 2023-03-23 \\
11.4167 & -18.4700 & -516.0000 & 0.1891 & 2.6172 & 10252 & 2023-03-23 \\
8.7333 & -23.1700 & -861.8000 & 0.3848 & 1.8850 & 10311 & 2024-03-23 \\
5.8333 & -23.1700 & -861.8000 & 0.4201 & 1.3986 & 10636 & 2024-03-23 \\
3.6950 & -23.1700 & -861.8000 & 0.3168 & 0.6117 & 10646 & 2024-03-23 \\
2.3117 & -23.1700 & -861.8000 & 0.1734 & 1.0498 & 10661 & 2024-03-23 \\
4.6650 & -23.1700 & -861.8000 & 0.1767 & 0.5593 & 10693 & 2024-03-23 \\
3.5400 & -23.1700 & -861.8000 & 0.2608 & 0.4584 & 10185 & 2024-03-23 \\
2.8467 & -23.1700 & -861.8000 & 0.3136 & 1.4474 & 10191 & 2024-03-23 \\
2.6667 & -23.1700 & -861.8000 & 0.1567 & 0.9965 & 10203 & 2024-03-23 \\
4.3333 & -23.1700 & -861.8000 & 0.1767 & 0.5593 & 10207 & 2024-03-23 \\
5.0033 & -23.1700 & -861.8000 & 0.1528 & 0.8374 & 10209 & 2024-03-23 \\
4.1000 & -18.1100 & -845.5000 & 0.4565 & 1.3780 & 10305 & 2021-11-03 \\
5.4833 & -18.1100 & -845.5000 & 0.3897 & 1.8850 & 10311 & 2021-11-03 \\
11.2000 & -18.1100 & -845.5000 & 0.3335 & 2.3597 & 10609 & 2021-11-03 \\
8.2500 & -18.1100 & -845.5000 & 0.1083 & 0.9648 & 10107 & 2021-11-03 \\
5.5167 & -18.1100 & -845.5000 & 0.2319 & 2.2405 & 10426 & 2021-11-03 \\
9.0833 & -18.1100 & -845.5000 & 0.2578 & 2.7939 & 10427 & 2021-11-03 \\
4.4100 & -18.1100 & -845.5000 & 0.4193 & 0.5914 & 10438 & 2021-11-03 \\
12.3733 & -18.1100 & -845.5000 & 0.3532 & 3.1458 & 10180 & 2021-11-03 \\
12.7950 & -18.1100 & -845.5000 & 0.3614 & 2.8969 & 10181 & 2021-11-03 \\
11.5933 & -18.1100 & -845.5000 & 0.3491 & 1.8959 & 10193 & 2021-11-03 \\
4.6333 & -18.1100 & -845.5000 & 0.2804 & 2.3549 & 10195 & 2021-11-03 \\
1.6350 & -18.1100 & -845.5000 & 0.1888 & 1.4198 & 10197 & 2021-11-03 \\
1.2267 & -18.1100 & -845.5000 & 0.1590 & 1.1070 & 10199 & 2021-11-03 \\
7.5883 & -18.1100 & -845.5000 & 0.1868 & 1.0500 & 10201 & 2021-11-03 \\
1.3533 & -18.1100 & -845.5000 & 0.1529 & 0.6097 & 10202 & 2021-11-03 \\
3.9917 & -18.1100 & -845.5000 & 0.1788 & 0.8957 & 10205 & 2021-11-03 \\
3.5467 & -18.1100 & -845.5000 & 0.1548 & 0.8374 & 10206 & 2021-11-03 \\
3.4250 & -18.1100 & -845.5000 & 0.1548 & 0.8374 & 10209 & 2021-11-03 \\
5.6267 & -18.1100 & -845.5000 & 0.1438 & 0.6415 & 10211 & 2021-11-03 \\
6.9050 & -20.4600 & -532.2000 & 0.1072 & 0.9648 & 10107 & 2024-08-11 \\
3.8133 & -20.4600 & -532.2000 & 0.4723 & 0.9662 & 10261 & 2024-08-11 \\
3.3517 & -20.4600 & -532.2000 & 0.4723 & 0.9662 & 10262 & 2024-08-11 \\
3.9083 & -20.4600 & -532.2000 & 0.4723 & 0.9662 & 10263 & 2024-08-11 \\
7.0167 & -20.4600 & -532.2000 & 0.1952 & 1.3052 & 10428 & 2024-08-11 \\
11.9527 & -20.4600 & -532.2000 & 0.7031 & 1.3587 & 10675 & 2024-08-11 \\
18.0287 & -20.4600 & -532.2000 & 0.5730 & 1.7879 & 10676 & 2024-08-11 \\
8.6603 & -20.4600 & -532.2000 & 0.5159 & 2.3574 & 10677 & 2024-08-11 \\
3.5917 & -20.4600 & -532.2000 & 0.1997 & 0.6881 & 10691 & 2024-08-11 \\
3.3317 & -20.4600 & -532.2000 & 0.1997 & 0.6881 & 10692 & 2024-08-11 \\
2.4567 & -20.4600 & -532.2000 & 0.1763 & 0.5593 & 10693 & 2024-08-11 \\
10.7833 & -20.4600 & -532.2000 & 0.3477 & 3.1458 & 10180 & 2024-08-11 \\
10.8750 & -20.4600 & -532.2000 & 0.3557 & 2.8969 & 10181 & 2024-08-11 \\
4.2750 & -20.4600 & -532.2000 & 0.3437 & 1.8959 & 10193 & 2024-08-11 \\
3.4800 & -20.4600 & -532.2000 & 0.3856 & 2.1945 & 10194 & 2024-08-11 \\
5.4317 & -20.4600 & -532.2000 & 0.1840 & 1.0500 & 10201 & 2024-08-11 \\
2.8500 & -20.4600 & -532.2000 & 0.1763 & 0.5593 & 10207 & 2024-08-11 \\
6.0533 & -20.4600 & -532.2000 & 0.1417 & 0.6415 & 10211 & 2024-08-11 \\
2.4467 & -20.4600 & -532.2000 & 0.1307 & 1.2388 & 10249 & 2024-08-11 \\
7.3017 & -20.8400 & -644.2000 & 0.2529 & 2.0985 & 10184 & 2025-04-15 \\
2.4500 & -20.8400 & -644.2000 & 0.3847 & 2.1945 & 10194 & 2025-04-15 \\
\end{longtblr}

%
%
\bibliography{ bibliography.bib }

@article{figueroa2025geomagnetic,
  title={{Geomagnetic disturbances and grid vulnerability: Correlating storm intensity with power system failures}},
  author={Figueroa, Mauro Gonz{\'a}lez and Acevedo, Daniel David Herrera and Porta, David Sierra},
  journal={PLoS One},
  volume={20},
  number={7},
  pages={e0327716},
  year={2025},
  publisher={Public Library of Science San Francisco, CA USA}
}

@techreport{kirkham2011geomagnetic,
  title={Geomagnetic storms and long-term impacts on power systems},
  author={Kirkham, Harold and Makarov, Yuri V and Dagle, Jeffery E and DeSteese, John G and Elizondo, Marcelo A and Diao, Ruisheng},
  year={2011},
  address={Richland, WA},
  institution={Pacific Northwest National Laboratory (PNNL), Richland, WA (United States)}
}

@article{malone2026mana,
  title={{The MANA magnetometer array, and magnetic observations across New Zealand from 2024}},
  author={Malone-Leigh, John and Rodger, Craig J and Hendry, Aaron T and Brundell, James and Mac Manus, Daniel and Petersen, Tanja and Evans, Lisa and Kruglyakov, Mikhail and Feng, Xinhu},
  journal={Space Weather},
  volume={24},
  number={2},
  pages={e2025SW004629},
  year={2026},
  publisher={Wiley Online Library}
}

@inproceedings{patel2016analysis,
  title={Analysis of geomagnetically induced current in transformer},
  author={Patel, JA and Mehta, RS and Rathod, SB and Patel, KJ and Rajput, VN and Pandya, KS},
  booktitle={{2016 International Conference on Electrical, Electronics, and Optimization Techniques (ICEEOT)}},
  pages={3909--3912},
  year={2016},
  organization={IEEE}
}

@article{patterson2024modelling,
  title={{Modelling electrified railway signalling misoperations during extreme space weather events in the UK}},
  author={Patterson, Cameron J and Wild, James A and Beggan, Ciar{\'a}n D and Richardson, Gemma S and Boteler, David H},
  journal={Scientific Reports},
  volume={14},
  number={1},
  pages={1583},
  year={2024},
  publisher={Nature Publishing Group UK London}
}

@misc{nerc2025reliability,
  title={{Reliability standards for the bulk electric systems of North America}},
  author={NERC},
  url={https://www.nerc.com/globalassets/standards/reliability-standards/rscompleteset.pdf},
  year={2025}
}

@article{seabold2010statsmodels,
  title={Statsmodels: econometric and statistical modeling with python.},
  author={Seabold, Skipper and Perktold, Josef and others},
  journal={scipy},
  volume={7},
  number={1},
  pages={92--96},
  year={2010}
}

@incollection{james2023statistical,
  title={Statistical learning},
  author={James, Gareth and Witten, Daniela and Hastie, Trevor and Tibshirani, Robert and Taylor, Jonathan},
  booktitle={An introduction to statistical learning: With applications in Python},
  year={2023},
  publisher={Springer}
}

@article{burnham2004multimodel,
  title={{Multimodel inference: understanding AIC and BIC in model selection}},
  author={Burnham, Kenneth P and Anderson, David R},
  journal={Sociological methods \& research},
  volume={33},
  number={2},
  pages={261--304},
  year={2004},
  publisher={Sage Publications Sage CA: Thousand Oaks, CA}
}

@article{santoso2024intensity,
  title={{The intensity of geomagnetic storms associated with the interplanetary magnetic field and solar wind parameters during Solar Cycle 24}},
  author={Santoso, Anwar and Sismanto, Sismanto and Priyatikanto, Rhorom and Hartantyo, Eddy and Martiningrum, Dyah R},
  journal={Earth and Planetary Physics},
  volume={9},
  number={2},
  pages={375--386},
  year={2024},
  publisher={Science China Press}
}

@article{burton1975empirical,
  title={{An empirical relationship between interplanetary conditions and Dst}},
  author={Burton, Rande K and McPherron, RL and Russell, CT},
  journal={Journal of geophysical research},
  volume={80},
  number={31},
  pages={4204--4214},
  year={1975},
  publisher={Wiley Online Library}
}

@article{ngwira2019introduction,
  title={An introduction to geomagnetically induced currents},
  author={Ngwira, Chigomezyo M and Pulkkinen, Antti A},
  journal={Geomagnetically induced currents from the sun to the power grid},
  pages={1--13},
  year={2019},
  publisher={Wiley Online Library}
}

@incollection{pirjola2012geomagnetically,
  title={Geomagnetically induced currents as ground effects of space weather},
  author={Pirjola, Risto},
  booktitle={Space science},
  volume={2},
  pages={27--44},
  year={2012},
  publisher={InTech Rijeka, Croatia}
}

@article{caraballo2025impact,
  title={{The impact of geomagnetically induced currents (GIC) on the Mexican power grid: Numerical modeling and observations from the 10 May 2024, geomagnetic storm}},
  author={Caraballo, R and Gonz{\'a}lez-Esparza, JA and Pacheco, CR and Corona-Romero, P and Arzate-Flores, JA and Castellanos-Velazco, CI},
  journal={Geophysical Research Letters},
  volume={52},
  number={4},
  pages={e2024GL112749},
  year={2025},
  publisher={Wiley Online Library}
}

@article{pulkkinen2001recordings,
  title={{Recordings and occurrence of geomagnetically induced currents in the Finnish natural gas pipeline network}},
  author={Pulkkinen, Antti and Viljanen, Ari and Pajunp{\"a}{\"a}, Kari and Pirjola, Risto},
  journal={Journal of Applied Geophysics},
  volume={48},
  number={4},
  pages={219--231},
  year={2001},
  publisher={Elsevier}
}

@article{boteler2013new,
  title={A new versatile method for modelling geomagnetic induction in pipelines},
  author={Boteler, DH},
  journal={Geophysical Journal International},
  volume={193},
  number={1},
  pages={98--109},
  year={2013},
  publisher={Oxford University Press}
}

@book{Baumjohann_2012,
    doi = {10.1142/p850},
    year = {2012},
    month = {Mar},
    title = {{Basic Space Plasma Physics}},
    publisher = {{I}mperial {C}ollege {P}ress},
    author = {Wolfgang Baumjohann and Rudolf A Treumann}
}

@article{boteler2014methodology,
  title={Methodology for simulation of geomagnetically induced currents in power systems},
  author={Boteler, David},
  journal={Journal of Space Weather and Space Climate},
  volume={4},
  pages={A21},
  year={2014},
  publisher={EDP Sciences}
}

@book{montgomery2019applied,
  title={Applied statistics and probability for engineers},
  author={Montgomery, Douglas C and Runger, George C},
  year={2019},
  publisher={John wiley \& sons}
}

@article{verma2025comprehensive,
  title={A comprehensive framework for residual analysis in regression and machine learning},
  author={Verma, Vibhu},
  journal={J. Inf. Syst. Eng. Manag},
  volume={10},
  pages={1--18},
  year={2025}
}

@misc{OMNI2020_1hr,
  doi = {10.48322/1SHR-HT18},
  url = {https://hapi-server.org/servers-dev/#server=CDAWebDev&dataset=OMNI2_H0_MRG1HR/source},
  author = {Papitashvili,  Natalia E. and King,  Joseph H.},
  title = {{OMNI Hourly Data Set [Kp, Dst]}},
  publisher = {NASA Space Physics Data Facility},
  year = {2020},
  copyright = {Creative Commons Zero v1.0 Universal}
}

@misc{OMNI2020_1min,
  doi = {10.48322/45BB-8792},
  url = {https://hapi-server.org/servers-dev/#server=CDAWebDev&dataset=OMNI_HRO2_1MIN/source},
  author = {Papitashvili,  Natalia E. and King,  Joseph H.},
  title = {{OMNI 1-min Data Set [AL, AU, SYM-H, Temp., Mag. Mach, Vx, By, Bz]}},
  publisher = {NASA Space Physics Data Facility},
  year = {2020},
  copyright = {Creative Commons Zero v1.0 Universal}
}

@article{Weigel2021,
  title = {{HAPI: An API Standard for Accessing Heliophysics Time Series Data}},
  volume = {126},
  ISSN = {2169-9402},
  DOI = {10.1029/2021ja029534},
  number = {12},
  journal = {Journal of Geophysical Research: Space Physics},
  publisher = {American Geophysical Union (AGU)},
  author = {Weigel,  Robert S. and Vandegriff,  Jon and Faden,  Jeremy and King,  Todd and Roberts,  D Aaron and Harris,  Bernard and Candey,  Robert and Lal,  Nand and Boardsen,  Scott and Lindholm,  Chris and Lindholm,  Doug and Baltzer,  Thomas and Brown,  Lawrence E. and Grimes,  Eric W. and Cecconi,  Baptiste and Génot,  Vincent and Renard,  Benjamin and Masson,  Arnaud and Martinez,  Beatriz},
  year = {2021}
}

@article{pulkkinen2008statistics,
  title={Statistics of extreme geomagnetically induced current events},
  author={Pulkkinen, A and Pirjola, R and Viljanen, A},
  journal={Space Weather},
  volume={6},
  number={7},
  year={2008},
  publisher={Wiley Online Library}
}

@article{Pulkkinen25,
    author = {Pulkkinen, A. A. and Schuck, P. and Bernabeu, E. and Weigel, R. S. and Arritt, R.},
    title = {A novel approach for geoelectric field response scaling factors used in geomagnetic storm hazard assessments},
    journal = {{Space Weather}},
    year = {2025}
}

@ARTICLE{2020SciPy-NMeth,
  author  = {Virtanen, Pauli and Gommers, Ralf and Oliphant, Travis E. and
            Haberland, Matt and Reddy, Tyler and Cournapeau, David and
            Burovski, Evgeni and Peterson, Pearu and Weckesser, Warren and
            Bright, Jonathan and {van der Walt}, St{\'e}fan J. and
            Brett, Matthew and Wilson, Joshua and Millman, K. Jarrod and
            Mayorov, Nikolay and Nelson, Andrew R. J. and Jones, Eric and
            Kern, Robert and Larson, Eric and Carey, C J and
            Polat, {\.I}lhan and Feng, Yu and Moore, Eric W. and
            {VanderPlas}, Jake and Laxalde, Denis and Perktold, Josef and
            Cimrman, Robert and Henriksen, Ian and Quintero, E. A. and
            Harris, Charles R. and Archibald, Anne M. and
            Ribeiro, Ant{\^o}nio H. and Pedregosa, Fabian and
            {van Mulbregt}, Paul and {SciPy 1.0 Contributors}},
  title   = {{{SciPy} 1.0: Fundamental Algorithms for Scientific
            Computing in Python}},
  journal = {Nature Methods},
  year    = {2020},
  volume  = {17},
  pages   = {261--272},
  adsurl  = {https://rdcu.be/b08Wh},
  doi     = {10.1038/s41592-019-0686-2},
}

@article{si2022impacts,
  title={Impacts of uncertain geomagnetic disturbances on transient power angle stability of DFIG integrated power system},
  author={Si, Yuan and Wang, Zezhong and Liu, Lianguang and Anvari-Moghaddam, Amjad},
  journal={IEEE Transactions on Industry Applications},
  volume={59},
  number={2},
  pages={2615--2625},
  year={2022},
  publisher={IEEE}
}

@article{matzka2021geomagnetic,
  title={{Geomagnetic Kp index}},
  author={Matzka, J{\"u}rgen and Bronkalla, Oliver and Tornow, Katrin and Elger, Kirsten and Stolle, Claudia},
  journal={GFZ Data Services},
  year={2021}
}

@article{cade2005quantile,
  title={Quantile regression reveals hidden bias and uncertainty in habitat models},
  author={Cade, Brian S and Noon, Barry R and Flather, Curtis H},
  journal={Ecology},
  volume={86},
  number={3},
  pages={786--800},
  year={2005},
  publisher={Wiley Online Library}
}

@article{koenker2005quantile,
  title={Quantile regression},
  author={Koenker, Roger},
  journal={Econometric Society Monographs, Cambridge University Press, Cambridge},
  year={2005}
}

@article{wilkerson,
author = {Wilkerson, L. A. and Weigel, R. S. and Thomas, D. and Bor, D. and Oughton, E. J. and Gaunt, C. T. and Balch, C. C. and Wiltberger, M. J. and Pulkkinen, A.},
title = {{GIC-Related Observations During the May 2024 Geomagnetic Storm in the United States}},
journal = {Space Weather},
volume = {24},
number = {5},
doi = {10.1029/2025SW004758},
year = {2026}
}

@book{heckert2002handbook,
  title={Handbook 151: Nist/sematech e-handbook of statistical methods},
  author={Heckert, N Alan and Filliben, James J and Croarkin, C M and Hembree, B and Guthrie, William F and Tobias, P and Prinz, J},
  year={2002},
  publisher={N. Alan Heckert, James J. Filliben, C M. Croarkin, B Hembree, William F~…}
}

@article{boteler2024examination,
  title={An examination of geomagnetic induction in submarine cables},
  author={Boteler, David H and Chakraborty, Shibaji and Shi, Xueling and Hartinger, Michael D and Wang, Xuan},
  journal={Space weather},
  volume={22},
  number={2},
  pages={e2023SW003687},
  year={2024},
  publisher={Wiley Online Library}
}

@article{love2025impact,
  title={{The impact of the May 1921 superstorm on American telecommunication systems}},
  author={Love, Jeffrey J and Lucas, Greg M and Kelbert, Anna and Schnepf, Neesha R and Bedrosian, Paul A and McBride, Sara K and Oughton, Edward J},
  journal={Space Weather},
  volume={23},
  number={7},
  pages={e2025SW004563},
  year={2025},
  publisher={Wiley Online Library}
}

@article{oughton2024major,
  title={Major Space Weather Risks Identified via Coupled Physics-Engineering-Economic Modeling},
  author={Oughton, Edward J and Bor, Dennies K and Weigel, Robert and Gaunt, C Trevor and Dogan, Ridvan and Huang, Liling and Love, Jeffrey J and Wiltberger, Michael},
  journal={arXiv preprint arXiv:2412.18032},
  year={2024}
}

@misc{NERC_ERO,
    title={{NERC ERO Portal}}, 
    url={https://eroportal.nerc.net/}, 
    journal={NERC}, 
    author={NERC}, 
    year={2026}
}

@misc{Thomas_Github,
    url = {https://github.com/dthoma6/SWERVE_analysis},
    year = {2026},
    title = {{SWERVE} analysis},
    urldate = {Jul 2026},
    author = {Thomas, Dean}
}

\end{document}